# Asymmetric matter-wave solitons at nonlinear interfaces

Fangwei Ye, Yaroslav V. Kartashov, and Lluis Torner

*ICFO-Institut de Ciencies Fotoniques, and Universitat Politecnica de Catalunya,*

*Mediterranean Technology Park, 08860 Castelldefels (Barcelona), Spain*

We predict the existence and study the basic properties of strongly asymmetric matter wave solitons that form at the interface produced by regions with different inter-atomic interaction strengths in pancake Bose-Einstein condensates. We address several types of surface solitons featuring topologically complex structures, including vortex and dipole-mode solitons. We found that the soliton become significantly asymmetric for high number of particles in the condensate. Yet, we reveal that under suitable conditions, that we elucidate, even such strongly asymmetric dipole and vortex solitons can be dynamically stable over wide regions of their existence domains.

The exploration of the rich nonlinear properties of Bose-Einstein condensates (BECs) is of paramount importance to understand the complex physics behind collective inter-atomic interactions. Nonlinearity affords the formation of different types of self-sustained structures [1] and manifest itself in such phenomena as four-wave mixing [2], wave amplification [3], superradiant Rayleigh scattering [4], or formation of Faraday patterns [5], to mention just a few examples. Solitons are of primary interest in this context. Both, dark [6] and bright [7] solitons have been observed in effectively one-dimensional condensates, while in two-dimensional geometries vortex solitons were created with phase imprinting technique [8], by rotating the BEC cloud [9], and via decay of solitons of other types [10]. On the other hand, a variety of vortex structures have been addressed in condensates with repulsive inter-atomic interactions, even in the regime of weak, quasi-vanishing interactions. This includes vortex dipoles [11], globally linked vortex clusters [12], vortex necklaces [13] and vortex lattices [14]. Vortex solitons in two-component condensates were considered in [15]. It was shown that vortex solitons in condensates with attractive inter-atomic interactions might be stable in the presence of harmonic traps, when the number of particles does not exceed a threshold level [16]. Finally, BECs loaded in optical lattices might also form stable vortex soliton states [17].

Typically, matter wave solitons featuring topologically nontrivial structures are studied in geometries with either attractive or repulsive inter-atomic interactions. Nevertheless, as it is well-established today, the magnitude and the sign of the interactions, characterized by the atomic scattering length $a_s$, can be tuned via Feshbach resonances by an external magnetic [18] or optical [19] fields. Feshbach resonances make it possible to change the scattering length $a_s$ not only in time (thus, e.g., inducing a temporal periodic variation of the nonlinearity strength), but also in space. The latter is central to the concept we put forward in this paper, and can be realized by applying of inhomogeneous field across the condensate. Feshbach resonance management in time results in existence of solitons of new types and may be potentially employed to suppress collapse of two-dimensional condensates that unavoidably develops for attractive interactions in the absence of trapping potentials [20, 21]. Spatial modulation of the scattering length results in new soliton dynamics [22]. Also, it can be used to create spatial domains featuring different strengths of inter-atomic interactions. Such domains are able to support new types of solitons, commonly termed surface solitons in other areas of physics, localized at the very interface that separates the domains.

The starting point of our analysis is the time-dependent Gross-Pitaevskii equation for the complex wave-function $\Psi(x,y,z,t)$ that describes the evolution of a BEC atom cloud in the frame of the mean-field approximation [1]:

$$i\hbar \frac{\partial \Psi}{\partial t} = -\frac{\hbar^2}{2m}\Delta\Psi + \frac{1}{2}m[\omega_r^2(x^2+y^2)+\omega_z^2 z^2]\Psi + g|\Psi|^2\Psi, \qquad (1)$$

where $\Delta$ is the three-dimensional Laplacian, $\hbar$ is the Plank constant, $m$ is the atomic mass, and $g = 4\pi a_s \hbar^2/m$ is the nonlinear coefficient with $a_s$ being the s-wave scattering length. We assume a cylindrically symmetric harmonic trapping potential with radial $\omega_r$ and axial $\omega_z$ frequencies. A "pancake" BEC configuration may be achieved when the axial frequency $\omega_z$ substantially exceeds the radial frequency $\omega_r$ ($\omega_z \gg \omega_r$). Combinations of radial magnetic and tight optical traps can be also used (see, e.g. Ref. [23] for a discussion and experimental details of the realization of condensates with different dimensionalities). Here we address condensates evolving in the weakly nonlinear regime and assume that they are always in the ground state of the harmonic potential in the axial direction. Denoting the ground state as $u_0(z)$, which in the weakly nonlinear regime can be described by a Gaussian function, one can factorize the 3D wave-function as $\Psi(x,y,z,t) = \Phi(x,y,t)u_0(z)$. Substitution of this ansatz into Eq. (1) and averaging in the $z$ direction leads to a normalized two-dimensional Gross-Pitaevskii equation (see, e.g., [5, 16, 21], for further details):

$$i\frac{\partial q}{\partial \tau} = -\frac{1}{2}\left(\frac{\partial^2 q}{\partial \eta^2} + \frac{\partial^2 q}{\partial \zeta^2}\right) + \frac{1}{2}\Omega(\eta^2 + \zeta^2)q + \sigma|q|^2 q. \tag{2}$$

Here $\tau = t\omega_0$ is normalized time, $\eta = x/r_0$ and $\zeta = y/r_0$, where $r_0 = (\hbar/m\omega_0)^{1/2}$ is the characteristic transverse scale, $\Omega = (\omega_r/\omega_0)^2$, $\omega_0$ is the characteristic trapping frequency, $\sigma(\eta,\zeta) = a_s(\eta,\zeta)/a_0$, $a_0$ is the characteristic scattering length, $q = (2\pi)^{-1/4}(4\pi a_0\hbar/\omega_0 r_z m)^{1/2}\Phi$, and $r_z = (\hbar/m\omega_z)^{1/2}$ is the width of the condensate cloud along the $z$ direction. In our calculations, the values of the frequency $\omega_0$ and the scattering length $a_0$ for typical experimental settings. Hence, $\omega_0$ was set to 200 Hz, while $a_0 \sim 22$ nm as in typical sodium condensates. Letting $\Omega = 1$ and $\sigma = 1$ in Eq. (2) one has that real radial frequency $\omega_r = \omega_0$ and real scattering length $a_s = a_0$. In sodium condensates a Feshbach resonance can be realized in the external magnetic field of the order of $B_0 \sim 900$ G (the width of the resonance is $\Delta = 1$ G) [24]. In this case, the value of the s-wave scattering length can be found from the empirically derived expression $a_s = a_r[1 - \Delta(B - B_0)^{-1}]$, where $a_r$ is the "off-resonance" scattering length [24], different from $a_0$, thus yielding different values of $\sigma = a_s/a_0$. For this set of parameters, $\tau = 1$ corresponds to 5 ms of actual cloud evolution, while the characteristic transverse scale amounts to $r_0 \sim 3.5$ $\mu$m.

Equation (2) conserves the number of particles $U$ and the Hamiltonian $H$:

$$\begin{aligned}U &= \int\int_{-\infty}^{\infty}|q|^2 d\eta d\zeta,\\ H &= \frac{1}{2}\int\int_{-\infty}^{\infty}[|\partial q/\partial \eta|^2 + |\partial q/\partial \zeta|^2 - \sigma(\eta,\zeta)|q|^4 + \Omega(\eta^2 + \zeta^2)|q|^2]d\eta d\zeta.\end{aligned} \tag{3}$$

The actual number of atoms $N$ in the condensate is related to the above norm by the expression $N = Ur_z/[(8\pi)^{1/2}a_s]$. In the case of axial trapping frequency $\omega_z = 2000$ Hz and for sodium condensates with $a_s \sim 22$ nm one gets $r_z \approx 1\,\mu$m and $N \approx 9U$.

We assume that the nonlinear coefficient $\sigma(\eta,\zeta)$ is tuned to a sharp transition, in such a way that in the limit case one can set

$$\sigma(\eta,\zeta) = \begin{cases} \sigma_{\text{left}} & \text{for } \eta < 0, \\ \sigma_{\text{right}} & \text{for } \eta \geq 0. \end{cases} \qquad (4)$$

This yields a sharp nonlinear interface at $\eta = 0$. However, we verified that the main qualitative results of this paper hold also for smooth interfaces, as discussed below. Without loss of generality, we set $\sigma_{\text{right}} = 1$, which implies repulsive inter-atomic interactions at $\eta \geq 0$, and vary $\sigma_{\text{left}}$ so that the condition $\sigma_{\text{left}} \leq \sigma_{\text{right}}$ holds. Creation of such nonlinear interfaces requires varying the Feshbach resonance across the condensate by means of suitable spatially inhomogeneous magnetic or optical fields.

We searched for soliton solutions of Eq. (1) numerically with a standard relaxation method. Here we are primarily interested in soliton solutions that exhibit topologically complex internal structures, such as so-called multi-pole solitons which have the general form $q = w\exp(i\mu\tau)$, and vortex solitons which can be written as $q = (w_r + iw_i)\exp(i\mu\tau)$. Here $w_{r,i}(\eta,\zeta)$ and $w(\eta,\zeta)$ are real functions independent of the normalized time $\tau$, while $-\mu$ stands for the chemical potential. The topological winding number (or topological charge) of the vortex solitons is obtained as the circulation of the phase gradient $\arctan(w_i/w_r)$ around a phase singularity located in the vicinity of the vortex core.

Representative examples of surface vortex solitons profiles with unit charge are depicted in Fig. 1. In the case shown the inter-atomic interactions are set to be repulsive at both sides of the interface, so that confinement of the condensate cloud along the transverse plane is achieved due to the external trapping potential. In the absence of the nonlinear interface, the vortex soliton profiles are radially symmetric with the vortex core (i.e., the wavefront singularity) located exactly at $\eta,\zeta = 0$. A similar picture is encountered in the quasi-linear limit when $\sigma_{\text{left}} \neq \sigma_{\text{right}}$ when the number of particles in the vortex is small and thus weak inter-atomic interactions do not lead to substantial profile distortion. Nevertheless, increasing the number of particles results in progressively large distortions of the vortex soliton profile (compare Figs. 1(a) and 1(b)). This is because the density of atoms in the condensate concentrates more and more in the regions where repulsive interactions are weaker. This generates strongly asymmetric and noncanonical vortex shapes supported by the interface but that only slightly penetrate into the region $\eta > 0$. The asymmetry of the vortex shape is accompanied by a shift in the location of the vortex core, so that the wavefront singularity is shifted progressively in the negative direction of $\eta$-axis. This is readily visible in Fig. 1. Increasing the number of particles, accompanied by progressively increasing shape asymmetry, results also in an overall expansion of the vortex core in the transverse plane. This is due to the repulsive character of the inter-atomic interactions assumed in this case. For a fixed $U$, the vortex asymmetry becomes more and more pronounced by decreasing the nonlinear coefficient $\sigma_{\text{left}}$.

The salient stationary properties of the surface vortex solitons are summarized in Fig. 2. All such solitons bifurcate from the linear radially symmetric eigenmodes of the parabolic trapping potential. Hence, they exist below an upper cutoff for $\mu$. The cutoff $\mu_{\text{co}}$ decreases monotonically with $\Omega$ (Fig. 2(a)). Interestingly, for a fixed chemical potential $-\mu$ and radial confinement $\Omega$, the surface vortex solitons can exist only when the strength $\sigma_{\text{left}}$ of the inter-atomic interactions at $\eta \leq 0$ exceeds a minimal value $\sigma_{\min}$. At fixed radial confinement $\Omega$, the critical value $\sigma_{\min}$ rapidly decreases as $\mu \to \mu_{\text{co}}(\Omega)$ (see Fig. 2(b)) and asymptotically approaches zero for $\mu \to -\infty$. Therefore, closer to the cutoff vortex solitons can be found even for attractive inter-atomic interactions at $\eta \leq 0$. Such structures contain a small number of particles and evolve almost in the linear regime. Notice that in spite of the attractive inter-atomic interactions, collapse can not develop for such solitons because the number of particles that they contain is far below the critical value. The extent of such structures in the transverse plane is determined mostly by the frequency of the parabolic potential. A similar picture is encountered when one fixes the chemical potential and varies $\Omega$, as illustrated in Fig. 2(c). In this case one finds that the critical value $\sigma_{\min} \to 0$ in weakly confining external potentials with $\Omega \to 0$. The number of particles $U$ is found to be a monotonically decreasing function of $\mu$ (Fig. 2(d)).

In view of the intrinsic asymmetry exhibited by the surface vortex solitons, one key issue to be elucidated is the dynamical stability of strongly asymmetric states. To this end, we performed detailed direct numerical integrations of Eq. (1) with input conditions $q|_{\tau=0} = (w_r + iw_i)(1+\rho)$, where $\rho$ stands for the profile of a noisy or a regular small perturbation $[\rho(\eta,\zeta) \ll w_{r,i}(\eta,\zeta)]$. Comprehensive simulations revealed that despite the strong shape asymmetry, the surface vortex solitons are stable in wide regions of their existence domain. Specifically, vortex solitons at the interface with $\sigma_{\text{left}} = 0.5$ are found to be unstable in a narrow interval around the approximate bounds $\mu \in [-1.55, -1.37]$ and stable in the rest of their existence domain, including the interval $\mu \in (-1.37, -0.89]$ adjacent to the linear cutoff and in the strongly nonlinear regime $\mu \to -\infty$. In terms of $U$ vortex solitons at $\sigma_{\text{left}} = 0.5$ are stable in the approximate ranges: $U > 34$ and $U < 22$, while at $\sigma_{\text{left}} = 0$ they are stable in the approximate ranges: $U > 283$ and $U < 6$. Using the estimate $N \approx 9U$ for total number of atoms derived above for sodium condensates and the fact that only condensates containing a large enough number of atoms, e.g., at least $N \sim 100$, are readily accessible experimentally, we conclude that *the strongly asymmetric vortex solitons* belonging to the left stability domains in Fig. 2(d), with correspondingly large enough $N$, *are the most relevant in practice*. Vortex solitons that belong to these stable regions retain their initial shapes over indefinitely long time intervals even in the presence of considerable input random perturbations, as it is clearly visible in Figs. 3(a) and 3(c). In contrast, the unstable solutions quickly reshape and dynamically evolve under the action of perturbations. For example, because of the action of the external trapping potential, solutions belonging to the left edge of the instability domain may undergo quasi-periodic splitting into pairs of solitons that subsequently recombine back. Notice that the width of the instability domain is not a monotonic function of $\sigma_{\text{left}}$. In particular, it shrinks completely when $\sigma_{\text{left}} \to \sigma_{\text{right}}$.

It is worth pointing out that surface vortex solitons can be also found for strong attractive interactions $\sigma_{\text{left}}, \sigma_{\text{right}} \sim -1$. Those are well localized even in the absence of an additional trapping potential. Nevertheless, such soliton solutions are highly prone to azimuthal modulational instabilities, so that they self-break apart into several fragments that either decay or collapse. Thus, here we do not study further such solutions.

We now turn to the properties of dipole surface solitons. Such solitons can be intuitively viewed as a nonlinear combination of two out-of-phase fundamental solitons that are glued together by the external trapping potential. Dipole solitons are characterized by the presence of a nodal line, where the modulus of the wave-function vanishes completely. In the absence of the nonlinear interface the nodal line is straight and poles are identical. In contrast, when $\sigma_{\text{left}} \neq \sigma_{\text{right}}$ the nodal line may become curved and poles acquire strong asymmetries. In principle, two different types of surface dipole solitons can be found: solitons whose poles reside in the regions $\zeta > 0$ and $\zeta < 0$ with straight nodal line perpendicular to the interface, and solitons whose poles carry a different fraction of total soliton norm and whose nodal lines are curved. Here we concentrate on the latter type (see Fig. 4 for representative examples of profiles). As in the case of surface vortices, dipole solitons exhibit almost canonical shapes at $U \to 0$ but become strongly asymmetric with increase of $U$. We found that dipole surface solitons bifurcate from the corresponding linear eigenmodes of the trapping potential and exhibit a cutoff for existence in $\mu$ identical to that for vortex surface solitons. Notice that at $\sigma_{\text{left}} \neq \sigma_{\text{right}}$ the distance between the nodal line and the interface increases with $U$. An increase of the number of particles causes an expansion of the dipole soliton in the transverse plane, accompanied by the appearance of multihumped soliton structures clearly observable in Fig. 4(b).

We found that domains of existence for surface dipole solitons are similar to those of vortex solitons. Thus, at fixed $\mu$ and $\Omega$ dipole solitons can exist only if $\sigma_{\text{left}}$ is above a minimal value $\sigma_{\min}$. Such minimal value decreases rapidly (and becomes negative) as $\mu$ approaches the cutoff $\mu_{\text{co}}(\Omega)$ for soliton existence. The dependencies $U(b)$ for different nonlinearity and trapping strengths are shown in Fig. 5. The number of particles $U$ in the dipole solitons decreases monotonically with $\mu$. Extensive numerical simulations of Eq. (1) with perturbed input conditions $q|_{\tau=0} = w(1+\rho)$ have revealed that asymmetric surface dipole solitons can be also stable in the region adjacent to the linear cutoff (see black curves in Figs. 5(a) and 5(b)). Importantly, while the width of stability domain in terms of $\mu$ only slightly changes with $\sigma_{\text{left}}$, the maximal number of particles $U$ in stable soliton dramatically increases for small values of $\sigma_{\text{left}} \sim 0.1$. This means that the interface plays a strong stabilizing action for the dipole solitons. Thus, we found that without the interface ($\sigma_{\text{left}} = \sigma_{\text{right}}$) dipole solitons can be stable only for $U < 20$, while, e.g., with $\sigma_{\text{left}} = 0$ the stability domain extends up to $U \approx 393$. This means that in currently available experimental settings, without interface dipole solitons are mathematically stable only for unrealistically small number of particles. Estimates may vary in specially prepared condensates, making these low-norm dipole solitons experimentally relevant. In any case, our results show that, in contrast, dipole solitons are readily observable at interfaces with $\sigma_{\text{left}} \to 0$, where they can contain up to $4 \times 10^3$ particles. Notice that in all cases increasing the asymmetry of the dipoles by increasing $U$ results in their destabilization. Solitons belonging to stable domains retain their structure over huge time intervals, while their unstable counterparts are quickly destroyed under the action of perturbations (Fig. 6). In the case of dipole solitons we found both exponential and oscillatory instabilities, while in the case of surface vortices we only found oscillatory instabilities.

Experimentally, the transition between regions with different strengths of inter-atomic interactions can not be made indefinitely sharp. To study the impact of the width of transition area on our predictions, we studied the case of smooth transitions, with the form $\sigma(\eta,\zeta) = (\sigma_{\text{right}} + \sigma_{\text{left}})/2 + (\sigma_{\text{right}} - \sigma_{\text{left}})\tanh(\eta/w)/2$ for the nonlinear coefficient in Eq. (2). Here the parameter $w$ sets the width of the transition area between regions with different nonlinearity strengths. Thus, $w \sim 1$ corresponds to a transition area of the order of a few microns, while in the limit $w \to 0$ one recovers the case of a step-like interface. Our comprehensive numerical simulations show that in the case of vortex solitons, increasing the width of the transition area results in a gradual shrinking of the instability domain (red curve in Fig. 2(d)). Also, the asymmetry of the vortex profile becomes less pronounced for smoother interfaces, whose width is comparable with the transverse extent of the vortex. In the case of dipole solitons, increasing the width of the transition region does not result in a substantial modification of stability domain. Experimental generation of the asymmetric soliton states may be achieved by the initial preparation of canonical (radially-symmetric) structures with screw phase dislocations, that were successfully demonstrated in several experiments on matter-wave vortices [8]. For example, topological phase dislocation may be imprinted in such radially-symmetric BECs by adiabatical inversion of the magnetic bias field in an Ioffe-Pritchard magnetic trap, or by phase imprinting. Feshbach resonance spatial management should then result in the asymmetrization of the vortex soliton shapes predicted in this paper.

Summarizing, we studied asymmetric matter wave solitons existing at novel type of condensate interfaces created by regions with different strengths of inter-atomic interactions. We studied in detail sharp interfaces, but we verified that the main qualitative results reported hold also for smooth interfaces. Such interfaces may be produced by properly tuning the Feshbach resonance with external magnetic or optical fields. The interfaces support various types of solitons with topologically complex and noncanonical internal structures, including vortex and dipole solitons. In spite of such strong asymmetries, we found that such solitons can be completely stable in wide regions of their existence domains, a property that facilitates their experimental observation.

This work was partially supported by the Generalitat de Catalunya, and by the Government of Spain through grant TEC2005-07815/MIC and the Ramon-y-Cajal program.

# Figure captions

Figure 1 (color online). Modulus of wave-function (left) and its phase (right) for surface vortex solitons at $\mu = -1.2$ (a) and $\mu = -2.5$ (b). In both cases $\Omega = 0.2$, $\sigma_{\text{left}} = 0.5$, and $\sigma_{\text{right}} = 1$. Vertical dashed lines show interface position.

Figure 2 (color online). (a) Cutoff for existence of vortex solitons vs $\Omega$. (b) Minimal nonlinearity strength vs $\mu$ at $\Omega = 0.2$. (c) Minimal nonlinearity strength vs $\Omega$ at $\mu = -1$. (d) Number of particles vs $\mu$ at $\Omega = 0.2$. Stable branches are shown in black, unstable branches are shown in red.

Figure 3 (color online). Dynamical evolution of perturbed surface vortex solitons at $\mu = -1.2$ (a), $-1.45$ (b), and $-2.5$ (c). Modulus of wave-function is shown for different moments of time. In all cases $\Omega = 0.2$, $\sigma_{\text{left}} = 0.5$, and $\sigma_{\text{right}} = 1$. Vertical dashed lines show the interface position.

Figure 4 (color online). Modulus of the wave-function for surface dipole solitons at $\mu = -1.5$ (a) and $\mu = -2.5$ (b). In both cases $\Omega = 0.2$, $\sigma_{\text{left}} = 0.1$, and $\sigma_{\text{right}} = 1$. Vertical dashed lines show the interface position.

Figure 5 (color online). Number of particles vs $\mu$ at $\Omega = 0.2$ (a) and $\sigma_{\text{left}} = 0.5$ (b). Stable branches are shown in black, unstable branches are shown in red.

Figure 6 (color online). Evolution of perturbed surface dipole solitons at $\mu = -1.5$ (a) and $-2.5$ (b). Modulus of wave-function is shown for different moments of time. In all cases $\Omega = 0.2$, $\sigma_{\text{left}} = 0.1$, and $\sigma_{\text{right}} = 1$. Vertical dashed lines show the interface position.

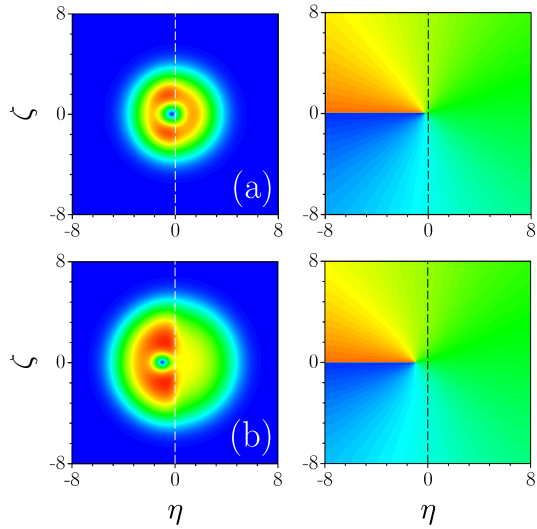

Figure 1 (color online). Modulus of wave-function (left) and its phase (right) for surface vortex solitons at $\mu = -1.2$ (a) and $\mu = -2.5$ (b). In both cases $\Omega = 0.2$, $\sigma_{\text{left}} = 0.5$, and $\sigma_{\text{right}} = 1$. Vertical dashed lines show interface position.

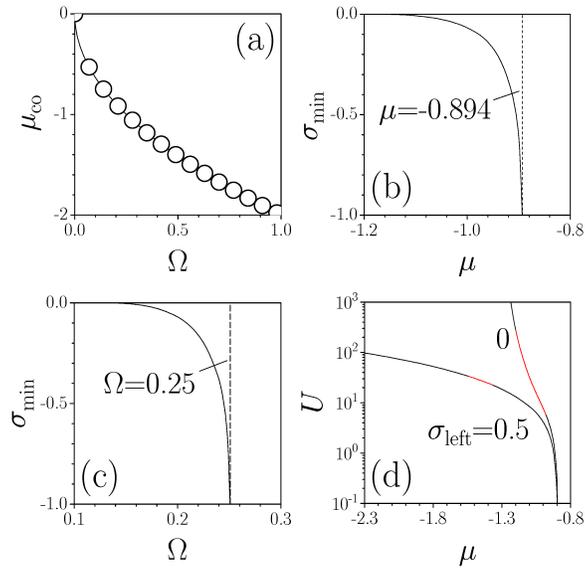

Figure 2 (color online). (a) Cutoff for existence of vortex solitons vs $\Omega$. (b) Minimal nonlinearity strength vs $\mu$ at $\Omega = 0.2$. (c) Minimal nonlinearity strength vs $\Omega$ at $\mu = -1$. (d) Number of particles vs $\mu$ at $\Omega = 0.2$. Stable branches are shown in black, unstable branches are shown in red.

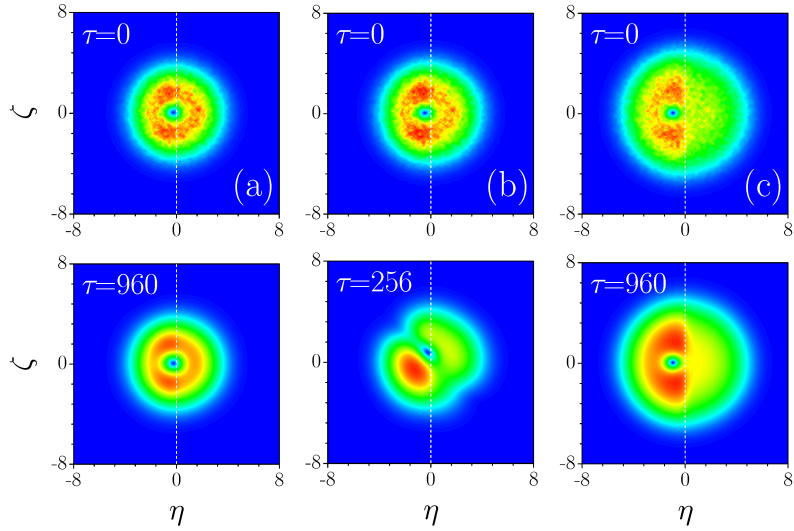

Figure 3 (color online). Dynamical evolution of perturbed surface vortex solitons at $\mu = -1.2$ (a), $-1.45$ (b), and $-2.5$ (c). Modulus of wave-function is shown for different moments of time. In all cases $\Omega = 0.2$, $\sigma_{\text{left}} = 0.5$, and $\sigma_{\text{right}} = 1$. Vertical dashed lines show the interface position.

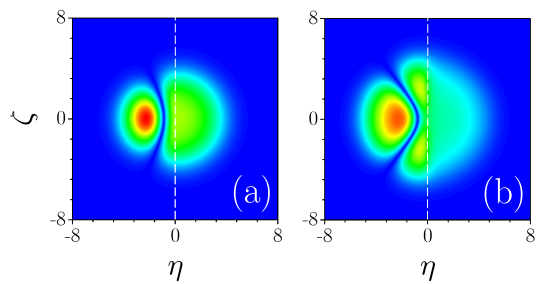

Figure 4 (color online). Modulus of the wave-function for surface dipole solitons at $\mu = -1.5$ (a) and $\mu = -2.5$ (b). In both cases $\Omega = 0.2$, $\sigma_{\text{left}} = 0.1$, and $\sigma_{\text{right}} = 1$. Vertical dashed lines show the interface position.

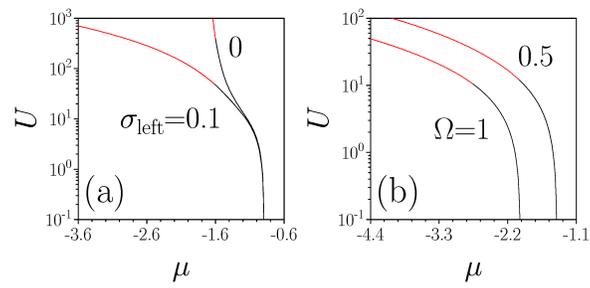

Figure 5 (color online). Number of particles vs $\mu$ at $\Omega = 0.2$ (a) and $\sigma_{\text{left}} = 0.5$ (b). Stable branches are shown in black, unstable branches are shown in red.

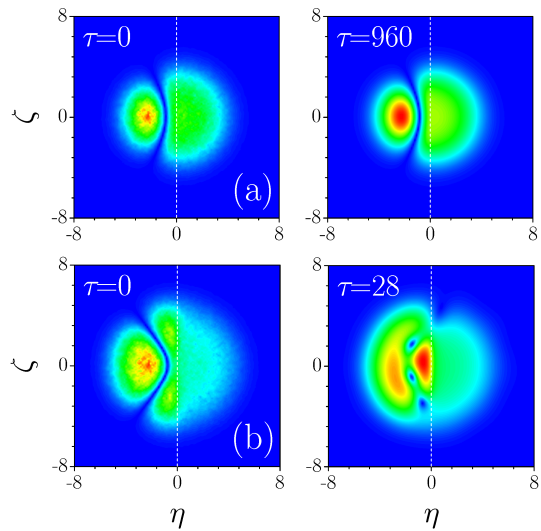

Figure 6 (color online). Evolution of perturbed surface dipole solitons at $\mu = -1.5$ (a) and $-2.5$ (b). Modulus of wave-function is shown for different moments of time. In all cases $\Omega = 0.2$, $\sigma_{\text{left}} = 0.1$, and $\sigma_{\text{right}} = 1$. Vertical dashed lines show the interface position.